  \documentclass[12pt]{article}
\input{epsf.sty}
\usepackage{aas_macros, natbib}
\begin{document}
\title{The Potential of Differential Astrometric Interferometry from the High Antarctic Plateau}

\author{James P. Lloyd, Ben R. Oppenheimer,  James R. Graham
} %

\date{}
\maketitle

\begin{center}{
Department of Astronomy, University of California, Berkeley USA\\
jpl@astron.berkeley.edu\\
}
\end{center}

\begin{abstract}
The low infrared background and high atmospheric transparency are the
principal advantages of Antarctic Plateau sites for astronomy.
However, the poor seeing (between one and three arcseconds) negates
much of the sensitivity improvements that the Antarctic atmosphere
offers, compared to mid-latitude sites such as Mauna Kea or Cerro
Paranal.  The seeing at mid-latitude sites, though smaller in amplitude, is
dominated by turbulence at altitudes of 10 to 20 km.  Over
the Antarctic plateau, virtually no high altitude turbulence is present in the
winter.  The mean square error for an astrometric measurement with a
dual-beam, differential astrometric interferometer in the very narrow
angle regime is proportional to the integral of $h^2 C_N^2(h)$.
Therefore, sites at which the turbulence occurs only at low altitudes
offer large gains in astrometric precision.  We show that a modest
interferometer at the South Pole can achieve 10 $\mu$as differential
astrometry 300 times faster than a comparable
interferometer at a good mid-latitude site, in median conditions.
Science programs that would benefit from such an instrument include
planet detection and orbit determination and astrometric observation
of stars microlensed by dark matter candidates.
\end{abstract}

{\bf Keywords:}
Instrumentation: Interferometers --- Techniques: Interferometeric ---
Astrometry --- Extrasolar Planets --- Galaxy: Halo, Stellar Content
\bigskip

\section{Introduction}

Properties of the atmosphere are the ultimate limit to astronomical
observations from ground-based sites.  Atmospheric transparency,
background and seeing (indicated by the full width of the point spread
function at half maximum, FWHM) are the parameters by which sites are
typically compared.  For observations in the optical and near
infrared, high mountain sites that are above most of the Earth's boundary
layer such as Mauna Kea and the Chilean Andes are superior because of
favorable values of these parameters.  Transparency and background are
parameters that are relatively simple to quantify and interpret.
However, there is much more to seeing than the FWHM of images taken
through the turbulent atmosphere.

The optical significance of turbulence in the atmosphere is due to
refractive index fluctuations, driven by temperature fluctuations.
These temperature fluctuations are usually described in terms of a
Kolmogorov model, with a temperature structure function as follows:
$$D_T(r) = \langle (T(x)-T(x+r))^2\rangle = C_T^2r^{2/3}.$$ 
Here, $T$ is the temperature, $x$ and $r$ are location variables, and
$C_T^2$ characterizes the amplitude of the turbulence.  These
temperature fluctuations cause refractive index ($N$) fluctuations
through the refractivity dependence of air on temperature, $C_N^2 =
(\delta N / \delta T)^2 C_T^2$.

The optical effect is characterized by the three-dimensional
refractivity power spectrum, which for Kolmogorov turbulence is
$$\Phi(\kappa,h) = 0.033 C_N^2(h) \kappa^{-11/3}.$$
Here, $\kappa$ is the spatial frequency of the turbulence, and
$C_N^2(h)$ describes the vertical profile, where $h$ is the height
above the observatory. $C_N^2(h)$ is typically complicated, with
multiple layers resulting from wind shear in the atmosphere.

The Fried parameter, $r_0$, relates the seeing to the equations above,
and is the scale over which the height-integrated phase fluctuation
at a given wavenumber, $k=2\pi/\lambda$, equals one radian RMS.

Given a $C_N^2(h)$ profile, the Fried parameter is
   $$r_0 = \left[0.423 k^2 sec(z) \int C_N^2(h) dh\right]^{-3/5},$$ 
and the FWHM of an
image is $0.98 \lambda/r_0$. $z$ is the zenith angle of
observation.  For the rest of this paper, we will refer all
calculations to the zenith.

Thus, the integrated amplitude of $C_N^2$ determines $r_0$ and
therefore the seeing.  The integrated seeing to ice level 
has proven to be a major drawback for
Antarctic telescopes, where the seeing is one to three arcseconds at
0.5 $\mu$m, a value more typical of poor, low-altitude sites than
superb sites such as Mauna Kea or Cerro Paranal, where the seeing is
routinely below an arcsecond.

However, other parameters, such as scintillation, isoplanatic angle,
and tilt anisoplanatism, which have paramount significance in certain
kinds of astronomical observations, depend on higher order moments of
the $C_N^2$ profile.  Specifically, tilt anisoplanatism is the
limiting error term for astrometric interferometry, and is a result
of the second moment of the $C_N^2$ profile.  The mean square error
for an astrometric measurement with a dual-beam, differential
astrometric interferometer in the very narrow-angle regime ($\theta h
<< B$) with long integrations, $t>>B/V$, is \citep{Shao92a}:
\begin{equation}   
\label{erroreqn}
\sigma_\delta^2 = 5.25 B^{-4/3}\theta^2 \int h^2 C_N^2(h) (Vt)^{-1}dh.
\end{equation}
In this equation, $B$ is the baseline of the interferometer, $\theta$
is the angular separation of the celestial objects, $V$ is the wind
speed as a function of height, and $t$ is total integration time of
the observation.  
This calculation assumes a Kolmogorov power spectrum
with no outer scale.
Due to the $h^2$ factor, this integral 
is completely dominated by
high altitude turbulence.  Thus, a site where the $C_N^2$ profile is
devoid of high altitude turbulence provides a substantial advantage.

The South Pole is such a site.  The $C_N^2(h)$ profile of the
atmosphere above the South Pole is fundamentally different from that of
any other characterised site \citep{Marks99}.  The unusual polar atmosphere is
entirely dominated by low altitude turbulence and is not affected by
the turbulence generating jet streams or trade winds nor the
high-altitude synoptic wind shear that creates turbulence at
mid-latitudes.

First, we explore the benefits of a South Pole interferometer
quantitatively, and then highlight two scientific projects for
such an interferometer.

\section{Benefits of a South Pole Interferometer}

The atomspheric turbulence at a site like Mauna Kea is dominated
by high altitude turbulence generated by tropopause
instability, jet streams, and wind shear between synoptic weather
systems.  
The lack of solar stratospheric heating at the South Pole 
during the polar winter results in a nearly adiabatic upper atmosphere, which
means that the optical effects of turbulence (due to temperature
fluctuations) are substantially reduced.
The tropopause is at least a factor of two lower and very weak.
Furthermore, there are no jet streams or trade winds at the 
South Pole, which are responsible for the large wind shears observed 
in atmospheric profiles at mid-latitude sites.

In order to compare the sites, we describe the $C_N^2(h)$ profile
in terms of a Hufnagel-Valley (HV) model \citep{Hardy, Roggemann}.
The HV model is an empirically fitted heuristic model, used
extensively in the context of theoretical calculations of light
propagation through the turbulent atmosphere.  The $C_N^2$ profile is
fitted by a series of terms: a planetary boundary layer term of the
form $C_N^2(h) = A_1 \exp(-h/H_1)$, a tropopause term of the form
$C_N^2(h) = A_2 h^{10}\exp(-h/H_2)$, and an individual layer term of
the form $C_N^2(h) = A_3 \exp(\frac{-(h-H_3)^2}{2d^2})$.  The
amplitudes, $A_n$, heights, $H_n$, and the thickness, $d$, of the
third component are all determined empirically.  The values of these
parameters used in our models of Mauna Kea and the South Pole are
indicated in Tables \ref{tab1} and \ref{tab2}.

Since the astrometric error is affected by the wind speed, we model
the atmospheric wind profile following \citet{Greenwood77}, but we
neglect the zenith angle and azimuth terms.  This model consists of a
constant ground layer term and a Gaussian profile tropopause wind
term:
$$V(h)=V_{\rm ground} + V_T\exp\left[-\left(\frac{h-H}{T}\right)^2\right].$$ 
For the Mauna Kea model, we use: $V_{\rm ground}=5$ m s$^{-1}$,
$V_T=25$ m s$^{-1}$, $H=12$ km, $T=5$ km.  For the South Pole model,
we use a higher ground wind speed, and a more moderate tropopause
speed at a lower altitude: $V_{\rm ground}=10$ m s$^{-1}$, $V_T=10$ m
s$^{-1}$, $H=7$ km, $T=3$ km.  The astrometric error is not strongly
sensitive to the details of the wind speed.

\citet{Marks99,Marks96} have measured the $C_N^2$ profile at the South
Pole with tower and balloon based microthermal sensors.  This profile
clearly shows that the vast majority of the turbulence is at altitudes
below 1 km.  Our HV model is fit to the median $C_N^2$ profile data of
\citet{Marks99}.  Integrating the HV model described in Table
\ref{tab2} and Figure \ref{fig2} results in seeing of 1.8 arcseconds
at 0.5 $\mu$m, typical for the site as measured by direct imaging
and differential image motion monitors.

We now use the model to assess the astrometric errors.  In order to
compare sites and our calculations with others in the literature, we
refer all calculations to an interferometer with a 100 m baseline,
reference stars separated by 1 arcminute and a 1 hour integration,
neglecting non-atmospheric noise sources.

\begin{figure}
\epsfxsize=6in
\centerline{\epsfbox{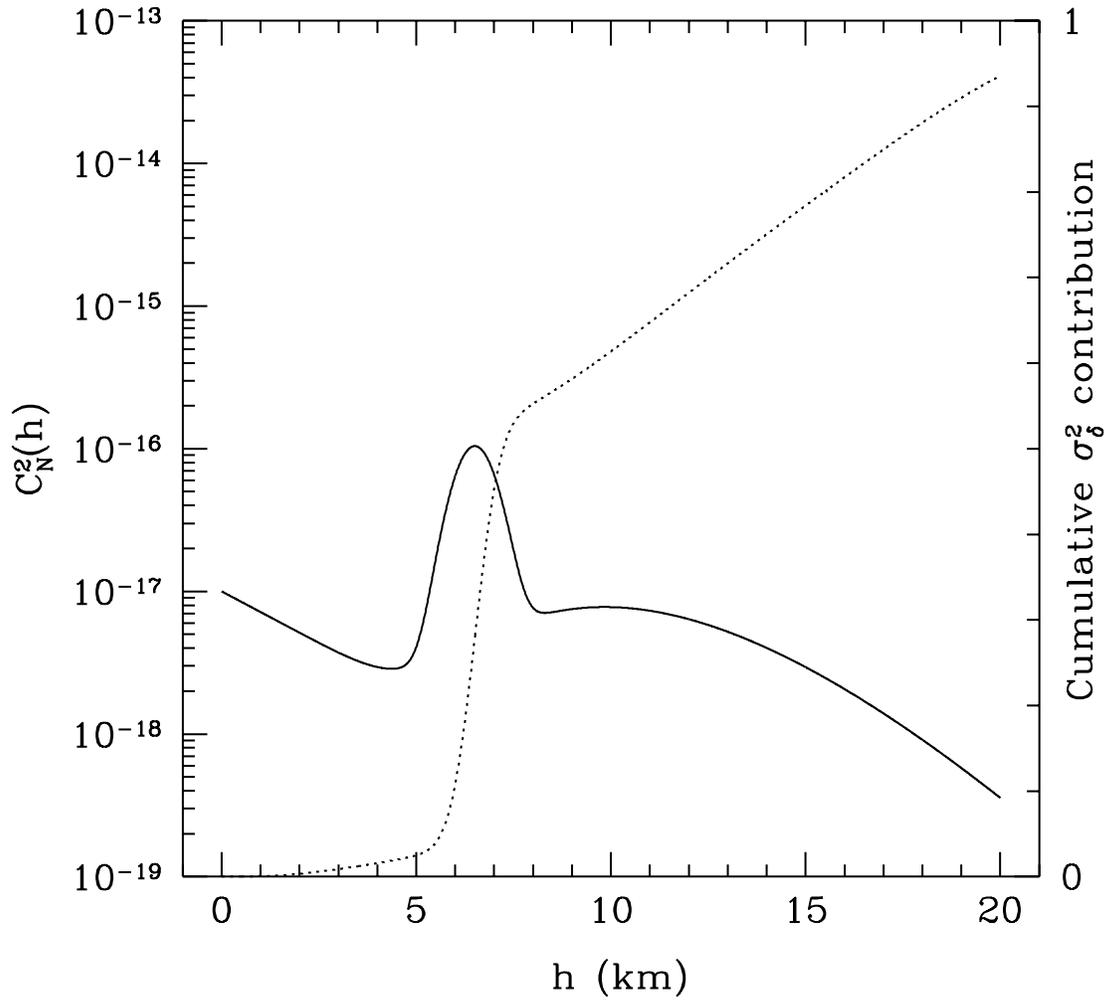}}
\caption{Mauna Kea Atmosphere Model \citep{Hardy}.
The solid line is the model $C_N^2(h)$ turbulent atmospheric 
profile.  The dotted line shows the cumulative contribution to
the mean square atmospheric astrometric error:
$\frac{\sigma_\delta^2(h)}{\sigma_\delta^2}
   = \frac{\int_0^h h'^2 V(h')^{-1} C_N^2(h') dh'}{\int_0^\infty h'^2 V(h')^{-1} C_N^2(h') dh'}$.\label{fig1}}
\end{figure}

\begin{figure}
\epsfxsize=6in
\centerline{\epsfbox{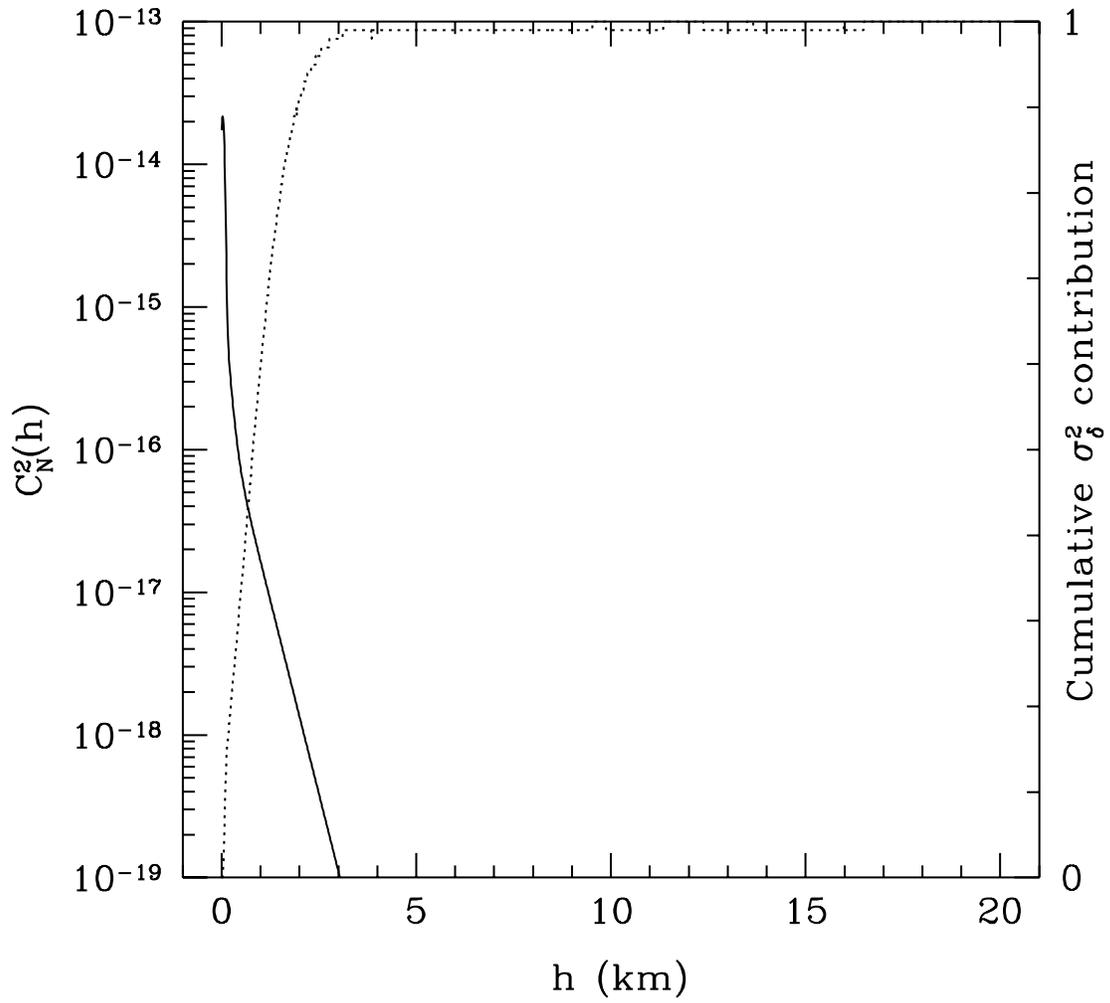}}
\caption{South Pole Atmosphere Model in Winter Adapted from
\citet{Marks99}.  The terms and notation are the same as in Figure \ref{fig1}. 
\label{fig2}}
\end{figure}

For the South Pole model, the astrometric error is $\sigma_\delta = 6$
$\mu$as.  We have also integrated a standard 0.5$^{\prime\prime}$
seeing Mauna Kea model \citep{Hardy} in Table \ref{tab1} and Figure
\ref{fig1}.  The astrometric error for the same interferometer
parameters is $\sigma_\delta = 100$ $\mu$as.  Another commonly used
reference atmosphere for Mauna Kea is that of \citet{Roddier90}, which
is the reference atmosphere used by \citet{Shao92a}.  Integrating this
0.3$^{\prime\prime}$ seeing profile we obtain an astrometric error of
$\sigma_\delta = 64$ $\mu$as, comparable to the \citet{Shao92a} value
of $\sigma_\delta = 60$ $\mu$as, for a 100 m baseline and a star
separation of 1 arcminute.   This mode of interferometry has been
demonstrated to achieve these levels of precision with the Mark III
and Palomar Testbed Interferometer \citep{Colavita94,Colavita99}.   

A measure of performance is the integration time required to achieve a
certain level of accuracy given the same interferometer configuration.
Equation \ref{erroreqn} indicates that the integration time is
inversely proportional to $\sigma_\delta^2$.  Therefore, these
calculations indicate that an astrometric interferometer at the South
Pole will achieve a given astrometric accuracy 300 times faster.
A 300-fold increase in the efficiency of observation enables a number
of science programs generally thought to be restricted to space-based
interferometers.

\begin{table}
\begin{center}
\caption{Mauna Kea Atmosphere Model 
\citep{Hardy}.\label{tab1}}
\begin{tabular}{lllll}
\\\hline\hline
Term            & Boundary Layer               & Tropopause                              & Strong Layer                           & Total \\
                & $A \exp(-h/H)$               & $A h^{10}\exp(-h/H)$                    & $A \exp(\frac{-(h-H)^2}{2d^2}$) & \\ \hline\hline
$A$             & $1\times10^{-17}$ m$^{-2/3}$ & $1.6\times10^{-53}$ m$^{-2/3}$m$^{-10}$ & $1\times10^{-16}$ m$^{-2/3}$    & \\ 
$H$ (m)           & 3000                        & 1000                                   & 6500                           & \\
$d$ (m)           &                              &                                         & 1000                           & \\
$r_0$ (cm)          & 65                         &  44                                   & 27                            & 20   \\
FWHM  & 0.15$^{\prime\prime}$   &  0.23$^{\prime\prime}$                            & 0.36$^{\prime\prime}$    & 0.5$^{\prime\prime}$ \\
$\sigma_\delta$ ($\mu$as) & 23                & 69                           & 69                   & 100  \\
\hline\hline
\end{tabular}
\end{center}
The contribution of individual terms are calculated, in addition to
the total.  The Fried parameter, $r_0$ is calculated at 0.5 $\mu$m.
The total $r_0$ is calculated as $r_0 = (\sum_{i}
{r_0}_i^{-5/3})^{-3/5}$.  FWHM = $0.98 \lambda/r_0$.  The calculated atmospheric astrometric
error, $\sigma_\delta$, is the root mean square error for a 100 m
baseline interferometer in 1 hour on stars separated by 1 arcminute,
neglecting all non-atmospheric noise sources.  $\sigma_\delta$ 
adds in quadrature.
\end{table}

\begin{table}
\begin{center}
\caption{South Pole Atmosphere Model in Winter Adapted from
\citet{Marks99}\label{tab2}}
\begin{tabular}{lllll}
\hline\hline
Term            & Lower Boundary Layer         & Upper Boundary Layer         & Strong Layer                           & Total \\
                & $A \exp(-h/H)$               & $A \exp(-h/H)$               & $A \exp(\frac{-(h-H)^2}{2d^2}$) & \\ \hline\hline
$A$             & $2\times10^{-15}$ m$^{-2/3}$ & $2\times10^{-15}$ m$^{-2/3}$ & $2\times10^{-14}$ m$^{-2/3}$    & \\ 
$H$ (m)         & 100                         & 400                         & 30                             & \\
$d$ (m)            &                              &                              & 40                             & \\
$r_0$ (cm)          & 21                         & 36                         & 6                             & 5.5  \\
FWHM& 0.48$^{\prime\prime}$    & 0.27$^{\prime\prime}$                  & 1.6$^{\prime\prime}$                      & 1.8$^{\prime\prime}$\\
$\sigma_\delta$ ($\mu$as)& 2                 & 5                & 2                   & 6 \\
\hline\hline
\end{tabular}
\end{center}
The terms and notation are the same as in Table \ref{tab1}.
\end{table}

This calculation is possibly optimistic.  A few balloon measurements
of $C_N^2$ \citep{Marks99} indicate that low-level high altitude
turbulence appears at times during the Antarctic Winter.  This
turbulence is enough to degrade the relative speed to a factor of
between 4 and 10 compared to Mauna Kea.  However, this is a comparison
of poor conditions at the South Pole to median or better at Mauna Kea.
Furthermore, the high altitude turbulence
measured by the balloon flights may be contaminated by wake turbulence
from the balloon itself \citep{Marks99}.  Although this hardly affects
the integrated seeing, the $h^2$ weighting for anisoplanatism strongly
emphasizes the high altitude turbulence.  Further studies are
warranted, particularly with SCIDAR measurements, which are most
sensitive to high altitude turbulence and are immune to wake
turbulence.

There are two independent, though circumstantial, pieces of evidence
that a large gain in astrometric error can be realised.
First, \citet{Marks99} calculates that the mean scintillation index,
$\sigma_I^2$, at the South Pole is four tenths of the value at Cerro
Paranal.  Since the scintillation index depends on a $h^{5/6}$ moment
of the $C_N^2$ profile, we can estimate a lower limit to the speed
improvement compared to Paranal as $0.4^{-12/5} = 9$.  

Second, the tilt anisoplanatism term is similar to the Strehl
anisoplanatism term that determines coherence over angular
separations.  Many people know the adage, ``Stars twinkle.  Planets
don't.''  This is generally true because planets are larger than the
isoplanatic angle at most locations on the Earth, so the
scintillations of different parts of the planet disk are
incoherent, and average out. If the isoplanatic
angle is larger than the disk of a planet, then the planet will
twinkle.  During the South Pole winters of 1995 and 1996, one of us
(JPL) saw Jupiter scintillate on occasion, indicating a isoplanatic
angle larger than 30 arcseconds in the visible, even at an elevation of 22
degrees above the horizon.
The mean conditions profile in Figure \ref{fig2} gives an isoplanatic angle of
33 arcsec at 0.5 $\mu$m at the zenith.   Since the isoplanatic angle
is proportional to airmass to the $8/5$ power, this naked eye
observation indicates that the zenith isoplanatic angle must
be as large as 2 arcmin under some conditions.
This is compelling evidence that the atmosphere
above the Antarctic Plateau is uniquely lacking high altitude
turbulence.  

The mechanism that generates the existing turbulence at the
South Pole is wind shear across the strongly inverted lower
atmosphere.  Such conditions are likely to exist anywhere
in the Antartic plateau, but the strength and height of the 
inversion, and windspeed vary markedly.  The height
of the inversion layer at the South Pole strongly correlates with 
wind speed, since the settling of cold air into a strong, shallow
inversion is disrupted by high wind.  Wind over the Antarctic plateau
is dominated by katabatic flow from the high points of the plateaus,
downhill towards the coast.  
Therefore, higher sites such as Dome C, where the mean windspeed
is only 2.6 m s$^{-1}$ are likely to show strong, shallow thermal
inversions, confining the turbulence generated to even lower
altitudes, which will offer even larger gains to astrometric
interferometry.  Likely to be the ulitimate site is the highest point on the
plateau, Dome A.  At these sites it is even possible that the
turbulent inverted layer may be shallow enough that it is feasible
to elevate the telescope above it, and realise the ``super seeing''
suggested by \citet{Gillingham93}.  Site testing programs
are underway to characterise these sites, including seeing 
monitoring programs.  Measurements of the $C_N^2$ profiles 
will be crucial to determination of the scientific programs
and instruments to be deployed at these sites.

\section{Science Applications}

The benefits shown in this calculation are limited to differential
astrometric observations.  Such observations have many applications.
We highlight two of them here, which are particularly promising with
regard to important astrophysical questions and well suited
to be conducted in the southern hemisphere.  We have also 
considered other configurations to take advantage of the unique
properties of the $C_N^2(h)$ profile.

The factor of 300 in speed achievable with an interferometer at the
South Pole result from the low altitude of the turbulence causing
common aberrations that can be removed with a differential measurement.
The same technique can be exploited in different ways with single
telescopes and adaptive optics or interferometers.  
The large
isoplanatic angle at the South Pole is attractive for adaptive
optics applications.  However, the small $r_0$ requires a very
bright guide stars and a high order AO system, even for a 
moderate sized telescope.  Similarly, a conventional interferometer 
measuring fringe visibility suffers from the poor seeing.  
A dual star interferometer is required to gain an advantage from the low
altitude of the turbulence with phase referencing, taking advantage of 
the coherence of the atmosphere across large angles.  Such an interferometer
can be used to measure fringe visibility on a faint source, while tracking
on a bright source.   
A generally useful interferometer in this mode
would require a large number of telescopes and long baselines to
allow bootstrapping the phase referencing to long baselines.
To measure relative fringe phase, however, only requires 
a single baseline, and it is disadvantageous to resolve the
source, so moderate baselines are desirable.  
We therefore conclude that astrometric interferometry is uniquely suited
to the site characteristics and logistical constraints in Antarctica.

\subsection{Planet Detection and Orbit Determination}

Astrometric observations with accuracies near a few $\mu$as permit the
detection and determination of the orbits of extrasolar planets.  For
reference, if a star like the Sun, lying at a distance of 10 pc, has a
Jupiter-like planet orbiting it, the star's position will be
perturbed by a maximum of 1 mas over the orbital period ($\sim 12$
years).  The signature of an Earth-like planet is approximately 1
$\mu$as.  Because the South Pole interferometer could, in principal,
conduct observations indefinitely, within 10 years one could not only
survey and determine the three dimensional orbital characteristics of
the planets discovered through radial velocity studies, but also
survey a substantial number of other stars within 30 pc for signatures
of extrasolar planets.  By comparison, the Space Interferometry
Mission (SIM), which has a limited lifetime of 5 years, will be unable
to observe stars over such long periods, limiting the types of planets
it will find.  It is necessary to fit the mass, semi-major axis, 
eccentricity, parallax and proper motion of the star in the astrometric
solution.  Accurate determination of all of these parameters 
with less than a complete orbit is not possible.  It will therefore
require significantly longer temporal baselines than SIM can 
provide to fully characterise the distribution of planets.

In comparison to the Keck Interferometer, the advantages of the South
Pole interferometer are the factor of 300 in speed of observation,
meaning that many more stars could be surveyed in a given time.
Alternatively, the substantially larger isoplanatic angle permits
observation of more stars.  For example, instead of speed, the
benefits result in a 300-fold increase in the
area of sky from which reference stars can be chosen.

\subsection{Astrometric Measurement of OGLE Microlensing Events}

There is a controversial body of evidence that suggests that a
substantial fraction of the dark matter in the Milky Way is in the
form of old white dwarfs.  These may have been detected by
microlensing experiments (MACHO; \citet{Alcock00}) and direct surveys
\citep{Opp01}.  However, the locations of the lensing masses remains
unmeasurable and is a major source of contention among various
researchers.  It is possible that none of the microlensing events
correspond to lenses in the Galaxy's halo.  Astrometry of the source
star during a microlensing event breaks the degeneracy in the solution
to the microlensing equations.  As a result, one can determine the
mass and distance to the microlens itself.  As shown by
\citet{Boden98}, astrometric accuracy of better than 10 $\mu$as
permits accurate determination of the lens parameters, although some
degeneracy remains concerning the transverse velocity of the lens.
Indeed, a lens placed at a distance of 8 kpc with the source star in
the Large Magellanic Cloud (LMC), produces an astrometric effect of
about 100 $\mu$as, readily detectable with a modest baseline 
South Pole interferometer.

Although the MACHO project \citep{Alcock00} is no longer searching for
microlenses, the OGLE-II project \citep{Ogle97} is currently engaged
in more sensitive observations of the LMC and their system for
broadcasting early detection of new events is almost operational.  The
LMC is easily observed from the South Pole.  With continuous
observation possible during the polar night, an interferometer
at the South Pole could resolve the remaining question concerning
the nature of the MACHO objects.

\section{Conclusions}

We have demonstrated that the unique structure of the atmosphere above
the high Antarctic plateau provides significant advantages to interferometric
astrometry.  The calculations described above motivate improved
measurements of the $C_N^2$ profile of the atmosphere 
with statistical coverage of varying conditions.  If such
measurements confirm our results, the construction of an
interferometer at the South Pole should be considered for the near
future.  The observations by such an instrument will have a
substantial impact on several important astrophysical problems.

\section*{Acknowledgments}

JPL has been supported in part by the National Science Foundation
Science and Technology Center for Adaptive Optics, managed by the
University of California at Santa Cruz under cooperative agreement
No. AST-9876783.  We thank the University of California--Berkeley
Center for Integrative Planetary Studies for a travel grant.  BRO is
supported by a Hubble Postdoctoral Research Fellowship (grant number
HF-01122.01-99A).  We thank Jean Vernin for providing electronic
versions of and discussions of South Pole $C_N^2$ profiles and Geoff
Marcy and Ben Lane for useful discussions.  
This work would not have been possible
without the dedicated work of Rodney Marks, who died at the South Pole
while wintering over.

\bibliographystyle{apj}
\bibliography{spam}

\end{document}